# Critical Analysis and Countermeasures Tactics, Techniques and Procedures (TTPs) that targeting civilians: A case study On Pegasus

1st Osama Hussien
Northumbria University
Department of Computer Science and Engineering
London, UK
ossama.akram@northumbria.ac.uk

2nd Usman Butt
Northumbria University
Department of Computer Science and Engineering
London, UK
usman.butt@northumbria.ac.uk

2nd Rejwan Bin Sulaiman
Northumbria University
Department of Computer Science and Engineering
London, UK
rejwan.binsulaiman@gmail.com

*Abstract*— Individuals, businesses, and governments all face additional difficulties because of the rise of sophisticated cyberattack attacks. This paper investigates the targeting of journalists and activists by the malware Pegasus. To gain a deeper understanding of the tactics utilized by cybercriminals and the vulnerabilities that facilitate their scope, this research looks on numerous occurrences and identifies recurring patterns in the strategies, methods, and practices employed. In this paper, a comprehensive analysis is conducted on the far-reaching consequences of these attacks for cybersecurity policy, encompassing the pressing need for enhanced threat intelligence sharing mechanisms, the implementation of more resilient incident response protocols, and the allocation of greater financial resources towards the advancement of cybersecurity research and development initiatives. The research also discusses how Pegasus will affect SCADA systems and critical infrastructure, and it describes some of the most important tactics that businesses may use to reduce the danger of cyberattacks and safeguard themselves against the 21st century's growing threats. The extent of Pegasus spyware, which can access various data and communications on mobile devices running iOS and Android potentially jeopardise the civil rights and privacy of journalists, activists, and political leaders throughout the world, was found to be worrying.

*Keywords—Pegasus spyware, Cyberattack tools, Cybersecurity policy, SCADA systems, Critical infrastructure, Privacy and civil liberties, Threat intelligence sharing, Incident response plans, Cybersecurity research and development, Tactics, techniques, and procedures of cybercriminals, Mobile device security, Journalists and activists as targets, Pegasus vulnerability analysis.*

I. INTRODUCTION

Cyberattacks targeting major businesses, human right advocate and journalists have increased in the past decade [1] and. [2]. The assaults damaged essential infrastructure, damaged finances, and damaged reputations. Attacks are becoming more frequent and sophisticated due to a multitude of variables, such as linked devices, cloud computing, and hackers' use of AI and machine learning. The Pegasus malware, used to covertly access and examine mobile devices in the modern day, stands out as one of the most noticeable and notable threats. State-sponsored actors and other groups use this tool to target human rights defenders, journalists, and activists, causing widespread disruption, making it critical for individuals and organizations to understand the latest hacking techniques and how to defend against them as the threat landscape evolves. The author will analyze the Pegasus spyware attack, including cybercriminal tactics and vulnerabilities, and briefly mention other common attacks in this research paper. In a comprehensive manner, the author will ultimately provide an extensive overview of crucial tactics and techniques that both individuals and organizations can effectively utilize to safeguard themselves from the constantly changing and advancing danger of cyber-attacks.

The selection of the Pegasus attacks and techniques was to provide a broad overview of different types of attacks and TTPs in the phishing attacks, while still focusing on the most significant and recent incidents such as Pegasus. The investigation endeavors to enhance the fortification of cyber safety by scrutinizing occurrences, deliberating on preventative measures, and comprehending how to safeguard against nascent menaces.

The aim of this paper is to provide an intricate and exhaustive account of the Pegasus spyware, expounding on its extensive and harmful impact on the basic rights of privacy and civil liberties, which are severely undermined by the sneaky and covert monitoring of journalists and activists by governments. Additionally, it seeks to emphasize the dangers and risks of Pegasus spyware, coupled with a thorough examination of the TACTICS, TECHNIQUES, AND PROCEDURES (TTPs) utilized by cyber offenders to carry out comparable attacks resembling Pegasus. Within the contents of this manuscript, the central point of interest pertains to the scrutinization and assessment of the methodologies utilized in the execution of contemporary and renowned cyber assaults, accompanied by an all-inclusive discourse on the most efficacious tactics and precautionary actions that can be enforced to minimize the probability of forthcoming attacks.

II. LITRETURE REVIEW

*A. Section remarks*

This section examines recent attacks/techniques using Pegasus. Additionally, it offers a critical evaluation of the nature of these attacks, the methods employed, and the possible effects on both enterprises and people. The study also examines potential defences against such attacks, as well as the analysis and reflections that might be made in response to them.

## B. Introductionn and History

Pegasus is spyware developed by the Israeli occupation cyber-arms company NSO Group that can be covertly installed on mobile phones (and other devices) running most versions of iOS and Android [3]. Pegasus is able to exploit iOS versions up to 14.7, through a zero-click exploit [3], which means that it can infect a device without any user interaction [3]. Pegasus can access various data and functions on the infected device, such as contacts, messages, photos, microphone, camera and location [3].

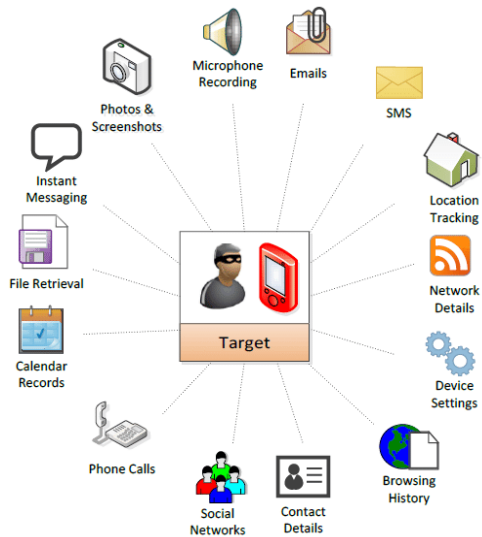

*Figure 1 Documentation reveals Pegasus can access various data from infected devices [4].*

One of the most notorious Pegasus spyware deployments involved the hacking of Amazon founder and Washington Post owner Jeff Bezos' smartphone. The Saudi Arabian government reportedly carried out the event as punishment for the Washington Post's negative coverage of the country, according to sources [5]. This incident demonstrated the potential of Pegasus to target influential figures and compromise their personal and professional data. However, Bezos was not the only victim of Pegasus. In 2019, WhatsApp discovered that Pegasus had been used to hack into the phones of multiple activists and journalists in India [6]. These attacks raised questions about the role of Pegasus in suppressing dissent and undermining democracy in India. Moreover, Pegasus has been implicated in violating the rights of human rights defenders in Palestine. In July 2021, an investigation by Front Line Defenders (FLD), a Dublin-based human rights group, found that the mobile phones of Palestinian rights defender and lawyer Salah Hammouri and five others were hacked using Pegasus [7]. This attack was particularly alarming given that Hammouri's Jerusalem residency status had already been revoked, raising concerns that the spyware was being used to further curtail his human rights work [7].

## C. Methodology

The first Pegasus version dates to 2016 was the CVE-2016-4657 with Apple's WebKit [8], that was referred to be an open-source browsing engine, so that 3rd programmers and sometimes even rivals can incorporate it into their own projects. For instance, the Nintendo Switch was vulnerable to this vulnerability since it utilized this WebKit in its native internet browser, which has been only intended for connecting to networks featuring captive portals [9] Due to a flaw in the JavaScript engine of WebKit, it can be vulnerable to this attack.

The threat actor makes use of the CVE-2016-4657 [8] flaw to get access on the system memory of Safari in WebKit. After that, a malware is installed on the target device that takes advantage of the kernel memory addressing leak caused by CVE-2016-4655 [10], Apple's approach for deserializing binary information lacks a size-checking function in one of the user-provided parameters that represents a 64-bit integer which allow this malware to proceed. The iOS kernel's address space configuration randomization option, which randomly generates the kernel image base through the boot loader before each boot, can thus be determined by the threat actor using this vulnerability [11].

To mitigate the Pegasus vulnerability, which allows the installation of a surveillance tool on the target's device, the malware initially deactivates code signing. Code signing ensures that code is secure and authentic. Disabling code signing is analogous to a doctor permitting a patient to ingest any substance, regardless of its safety. Another flaw, CVE-2016-4656 [12], allows the malware to reallocate previously freed memory from a string and insert a stack pivot into the NULL page in order to execute code in a privileged environment. The vulnerability enables root access by giving an attacker access to a shell [11].

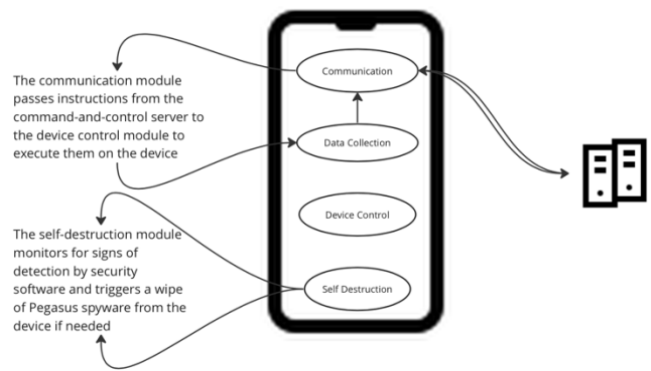

*Figure 2 Inner working and methodology of Pegasus*

From this section, and according to [18] it was concluded that Pegasus contain several key component as described In figure 2, starting with the Communication module, which is responsible for sending and receiving data between the device and the command-and-control server controlled by the attacker, the Data Collection Module is responsible for harvesting data from various sources on the device, such as messages, calls, photos, videos, location, passwords and apps, while the Device Control module is responsible for manipulating the device's functions, such as turning on or off the microphone or camera, recording audio or video, deleting files or apps, finally, the Self Destruction module is responsible for removing traces of Pegasus spyware from the device when instructed by the attacker or when detected by security software [18].

Figure 3 sums up the details of how the Pegasus uses Zero Days vulnerabilities to attack any vulnerable device [18]

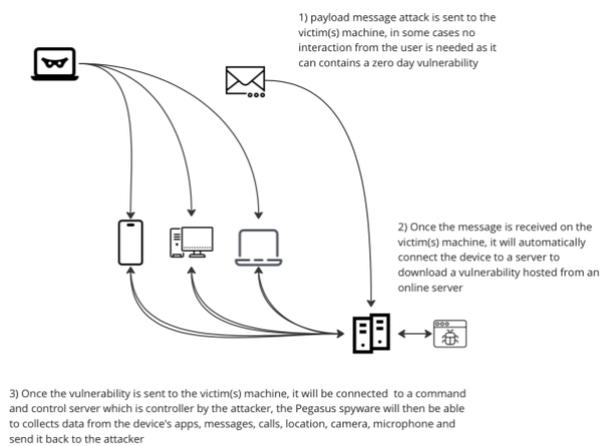

*Figure 3 Pegasus exploitation process*

D. General Impact

The Pegasus vulnerability poses a grave risk to privacy and civil liberties, as it enables governments and other hostile actors to infiltrate individuals and organizations without their awareness or consent [13]. The capability to remotely access confidential data and communication on a mobile device constitutes a severe breach of privacy, as well as a potential instrument for surveillance and censorship. A major concern regarding the use of Pegasus is that it can be employed to target journalists, activists, and other individuals who are opposed to government policies or involved in human rights work [13]. By tracking their communication and activities, governments can monitor and intimidate these individuals, possibly resulting in the silencing of free speech and the limitation of civil liberties [13].

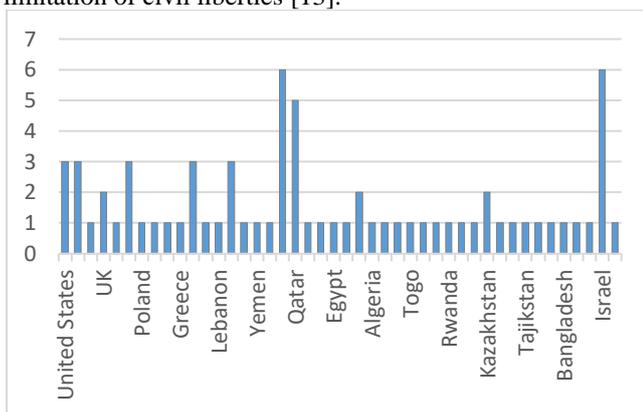

*Figure 4 Suspected Pegasus usage intensity in different company [4]*

The global reach and impact of Pegasus spyware on human rights was revealed by a report based on DNS cache probing on domain names extracted from command and control (C&C) servers. The report [4] found that at least 45 countries were suspected of having Pegasus infections, operated by at least 33 likely NSO customers, including governments, intelligence agencies, and law enforcement agencies. The report also noted that Pegasus had been used to target journalists, activists, and opposition politicians in various countries, raising concerns about privacy and civil liberties violations. The report called for greater transparency from companies like NSO Group, the Israeli company that developed Pegasus, and for greater regulation of the surveillance industry to prevent abuses. However, the Pegasus vulnerability also poses challenges for governments in regulating and monitoring the development and use of cyberweapons. While some believe that it is the duty for governments to safeguard their population from cyber dangers, some pointing out that the widespread use of technologies like Pegasus can weaken democracy and may ultimately be ineffective in the battle against terrorism and other criminal behaviour [14]. Therefore, when dealing with cyberweapons like Pegasus, an appropriate approach must be used that takes into account security issues and human rights concerns.

Tight rules for the creation and use of those tools are required in light of the major concerns highlighted by the use of Pegasus spyware over the increasingly sophisticated nature of cyberattacks. The NSO Group has been under controversy for providing its spyware to nations having an established history of violating human rights, and requests for stricter export controls have increased in recent years which leaded that its on the US Black List now [15], the widespread use of Pegasus and other advanced spyware highlights the need for greater regulation and oversight of the cybersecurity industry to prevent the abuse of these tools by authoritarian regimes and other malicious actors.
More openness and responsibility in the creation and application of cyberweapons are required to solve this issues. Authorities need to be held accountable for any kind of misuses of there authority or invasions of privacy that occur as a result of using this technologies and spywares and they should be forced to report how they use them there ought to be increased efforts to support and safeguard human rights and privacy, especially the improvement of encryption as well as additional privacy-enhancing technology [16].

E. SCADA Impact

Another serious threat caused by Pegasus is the operational integrity and reliability of SCADA systems and critical infrastructure. By exploiting various vulnerabilities in the system, Pegasus spyware can access and manipulate data, commands, sensors, actuators, and other components of SCADA systems. This could result in loss of control, malfunctioning, damage, or shutdown of critical processes and equipment. For example, Pegasus spyware could alter the pressure or temperature readings of a gas pipeline or a nuclear reactor, causing leaks or explosions Alternatively, it could disrupt the power supply or communication networks of a transportation system or a hospital, affecting the safety and efficiency of these services [17] and with the power of being able to infiltrate devices with zero clicks, it can be catastrophic. It can also cause a significant financial implication. A cyberattack on these systems could cause direct costs such as repair expenses, fines, lawsuits, compensation claims, or ransom payments [17]. Moreover, it could cause indirect costs such as loss of revenue [17], reputation damage customer dissatisfaction, or competitive disadvantage. For instance, Pegasus spyware could steal confidential information or trade secrets from an industrial company or a utility provider [18], giving an advantage to its competitors or adversaries. Additionally, Pegasus spyware could expose sensitive data such as personal information, financial records or health records of customers or employees of these systems leading to identity theft or fraud [18].

Researchers at [19] highlights the challenge of zero-day vulnerabilities on SCADA Systems that can be exploited by malicious actors before they are patched. A recent example of such a vulnerability is CVE-2021-30860 [20], which was used by Pegasus spyware to infect iOS devices without any user interaction. The paper's analysis and suggestions are also relevant and timely for cybersecurity and human rights in light of the Pegasus vulnerability and its misuse by authoritarian governments, the same researcher proposed a simulation and detection framework to protect SCADA systems from ransomware attacks, which can exploit zero-day vulnerabilities like Pegasus [21].

*F. Covid-19 and Hybrik/From Home Work Envirionment Impact*

According to [17], the high number of Internet of Things (IoT) devices linked to home networks is one of the reasons why they are vulnerable to Pegasus. According to [17], In the digital age, the average family has 10 IoT gadgets. The more of the internet of things devices there are, the less difficult it is for hackers to enter the network by breaking into them. Additionally, 59.7% of residences had routers that were susceptible to hacker assaults and password changes [18]. To prevent these threats, some governments and organizations have strict policies that forbid employees from using their own devices for work purposes [19] [20]. Other security measures include updating cyber hygiene practices, providing security awareness training, and revising cyber hygiene rules. However, these precautions may not be enough to stop advanced attacks like Pegasus [21].
Researchers at [17] also highlight other challenges and threats faced by remote workers who are working from home due to the global pandemic. Although they proposed some robust protocols for organizations to protect their remote workers and corporate networks from cyberattacks. However, these protocols may not be enough to prevent the Pegasus spyware, which can exploit zero-day vulnerabilities, remote workers who use their personal devices for work purposes may be at risk of this spyware and compromise their sensitive data and networks.

*G. Related and Similar attack vectors*

This recent attacks has highlighted the importance of maintaining updated security systems, replacing legacy systems, regular backups, security awareness training and implementing multi-factor authentication to guard against unauthorized access. This ransomware attack is one of many examples of how cyber threats are on the rise globally. Other notable examples includes The 2017 Equifax data breach [17] and the 2016 hack of the DNC [18] are other high-profile cyber attacks that have occurred in recent years. State-sponsored cyberattacks are also rising in regularity [19] and [20]. The governments of China and Russia was charged with funding cyber espionage against numerous agencies of the United States and commercial companies [21] and [22]. To guard from such attacks, organisations must make sure they adopt strict safety protocols, such as strict access controls, multi-factor authentication, and frequent security assessments.

## III. TACTICS, TECHNIQUES, AND PROCEDURES, AND COUNTER MEASURES

Cybersecurity is a critical concern in today's digital world, with cyberattacks becoming increasingly prevalent and sophisticated. Attackers penetrate networks, steal data, and create disruption using a range of attack vectors and TTP. It might be difficult and slow down the process of identifying an attacker's TTPs to retrieve security data from unstructured material [23]. The problem was addressed by several researchers using a variety of approaches, including a thorough evaluation of various Natural Language Processing (NLP) and machine learning techniques, most notably a data processing pipeline that classifies unstructured content into attackers' tactics and techniques by using a knowledge base of adversary TTPs [24], that makes it possible for textual data to be automatically and promptly extracted in order to extract crucial security information, supporting efficient threat detection and response.

One common TTP used by attackers is social engineering and phishing attacks [25], It entails coercing people into disclosing private information or allowing access to systems. Other TTPs used to obtain unauthorised access to systems and data include spear-phishing, malware, and brute-force assaults. Attackers also employ strategies like ransomware to encrypt private information and demand payment from their targets. In order to create efficient defences and safeguard against cyberattacks, it is essential to understand the TTPs employed by attackers. To protect themselves from these risks, organisations must develop effective security measures and keep up with the most recent TTPs.

*Table 1 types of breaches or attacks in 2022, among the organizations [24]*

| Attack Type/Organization Type | Phishing | Impersonation | General Malware | Denial Of Service | Online Banking Attack | Organization Account Takeover | Ransomware | Outsider unauthorized access | Unauthorized listening to video conference/ or Instant Messages | Insider unauthorized access |
|---|---|---|---|---|---|---|---|---|---|---|
| Organization | 83 | 27 | 12 | 10 | 8 | 8 | 4 | 2 | 1 | 1 |
| Charities | 87 | 26 | 11 | 2 | 6 | 6 | 4 | 2 | 3 | 1 |

To defend against these attacks, individuals and organizations must also implement best practices for cybersecurity, such as regularly updating software and hardware to address vulnerabilities, using strong passwords and limiting access to sensitive information [26]. Security protocols such as firewalls and intrusion detection and prevention systems, endpoint protection software, encryption technologies and Security Information and Event Management (SIEM) tools can be used to aggregate and analyse security events across the network, providing greater visibility into potential threats and help can mitigate any unauthorized access to systems and data [27]. Regular security assessments and employee training on recognizing and avoiding cyberattacks can also significantly reduce the risk of successful attacks.

The Pegasus attacks are a prime example of how state-sponsored actors can use sophisticated spyware to compromise the security of mobile devices. Keeping mobile devices updated with the latest security patches, using strong passwords, and being cautious of unknown links and attachments are important countermeasures [28].

Common methods used in cyberattacks include phishing, malware, social engineering, and credential stuffing. These methods rely on taking advantage of flaws in software, hardware, or human factors to access systems and data which are easier to execute with less technical knowledge thanks to the adoption of tools like Metasploit and the Burp suite [29]. Organizations must educate their employees about these techniques and regularly conduct vulnerability assessments and penetration testing to identify potential weaknesses.

The success of cyber-attacks can be attributed to the existence of unremedied software, feeble passcodes, and unguarded hardware susceptibilities. To effectively shield their networks from looming dangers and ensure the safeguarding of confidential data, it is of utmost importance for enterprises to take a pre-emptive approach to security by enforcing rigorous entry restrictions, carrying out regular security inspections, and utilizing state-of-the-art security mechanisms. The malevolent acts of cyber criminals are a formidable menace, therefore it is imperative to undertake measures to safeguard oneself against them. By embracing a pre-emptive stance towards cybersecurity and keeping abreast of the most recent perils, people and institutions can greatly mitigate the likelihood of succumbing to digital assaults.

## IV. CONCLUSION

It is essential to create a comprehensive plan for defending against such threats as the complexity and severity of cyberattacks increase in order to effectively reduce the risks involved. Undoubtedly, the appearance of Pegasus is a particularly unsettling example of such attacks because it exploits a variety of vulnerabilities in iOS devices to infect and track its victims without their knowledge or consent. This paper emphasizes the importance of understanding the TTPs used by attackers, discussing vulnerabilities and countermeasures to provide valuable insights to organizations and individuals. Implementing best practices such as strong passwords, regular security assessments, and advanced security technologies can help mitigate the risk of exploitation of known vulnerabilities. Users must routinely update their devices nevertheless, and keep an eye out for any strange behaviour and it's important to note that It is crucial for governments and international organisations to control the use of cyberarms like Pegasus and make those responsible who misuse them for nefarious reasons accountable.

## V. ACKNOWLEDGMENTS

The authors would like to express their gratitude to {BLINDED CO-AUTHOR 1} and {BLINDED-CO AUTHOR 2} for their valuable contributions to this research. Their expertise, assistance, and support were instrumental in the successful completion of this study.

## VI. CONFLICT OF INTEREST

The authors declare no conflicts of interest related to the publication of this research article.